\documentclass[fleqn,11pt]{article}

\usepackage{latexsym,ifthen,epsfig,color}
\usepackage{amsmath,amssymb,amsthm}
\usepackage{float}

\newcommand{\join}{\text{\textcircled{{\footnotesize 1}}}}
\newcommand{\cojoin}{\text{\textcircled{{\footnotesize 0}}}}

\newcommand{\NP}{\ensuremath{\mathbb{NP}}}

\newtheorem{theorem}{Theorem}
\newtheorem{lemma}{Lemma}

\newtheorem{corollary}{Corollary}
\newtheorem{proposition}{Proposition}

\newtheorem{clai}{Claim}
\newtheorem{observation}{Observation}
\newtheorem{algo}{Algorithm}[section]
\newtheorem{proc}{Procedure}[section]

\begin{document}

\title{Dominating Induced Matchings in $S_{1,2,4}$-Free Graphs}

\author{
Andreas Brandst\"adt\footnote{Institut f\"ur Informatik,
Universit\"at Rostock, A.-Einstein-Str.\ 22, D-18051 Rostock, Germany,
{\texttt andreas.brandstaedt@uni-rostock.de}}
\and
Raffaele Mosca\footnote{Dipartimento di Economia, Universit\'a degli Studi ``G.\ D'Annunzio''
Pescara 65121, Italy.
{\texttt r.mosca@unich.it}}
}

\maketitle

\begin{abstract}
Let $G=(V,E)$ be a finite undirected graph without loops and multiple edges. A subset $M \subseteq E$ of edges is a {\em dominating induced matching} ({\em d.i.m.}) in $G$ if every edge in $E$ is intersected by exactly one edge of $M$. In particular, this means that $M$ is an induced matching, and every edge not in $M$ shares exactly one vertex with an edge in $M$. Clearly, not every graph has a d.i.m.

The \emph{Dominating Induced Matching} (\emph{DIM}) problem asks for the existence of a d.i.m.\ in $G$; this problem is also known as the \emph{Efficient Edge Domination} problem; it is the {\em Efficient Domination} problem for line graphs.

The DIM problem is \NP-complete in general, and even for very restricted graph classes such as planar bipartite graphs with maximum degree 3. However, DIM is solvable in polynomial time for claw-free (i.e., $S_{1,1,1}$-free) graphs, for $S_{1,2,3}$-free graphs, for $S_{2,2,2}$-free graphs as well as for $S_{2,2,3}$-free graphs, in linear time for $P_7$-free graphs, and in polynomial time for $P_8$-free graphs ($P_k$ is a special case of $S_{i,j,\ell}$). In a paper by Hertz, Lozin, Ries, Zamaraev and de Werra, it was conjectured that DIM is solvable in polynomial time for $S_{i,j,k}$-free graphs for every fixed $i,j,k$.   

In this paper, combining two distinct approaches, we solve it in polynomial time for $S_{1,2,4}$-free graphs which generalizes the $S_{1,2,3}$-free as well as the $P_7$-free case.
\end{abstract}

\noindent{\small\textbf{Keywords}:
dominating induced matching;
efficient edge domination;
$S_{1,2,4}$-free graphs;
polynomial time algorithm.
}

\section{Introduction}\label{sec:intro}

Let $G=(V,E)$ be a finite undirected graph. A vertex $v \in V$ {\em dominates} itself and its neighbors. A vertex subset $D \subseteq V$ is an {\em efficient dominating set} ({\em e.d.s.} for short) of $G$ if every vertex of $G$ is dominated by exactly one vertex in $D$.
The notion of efficient domination was introduced by Biggs \cite{Biggs1973} under the name {\em perfect code}.
The {\sc Efficient Domination} (ED) problem asks for the existence of an e.d.s.\ in a given graph $G$ (note that not every graph has an e.d.s.)

A set $M$ of edges in a graph $G$ is an \emph{efficient edge dominating set} (\emph{e.e.d.s.} for short) of $G$ if and only if it is an e.d.s.\ in its line graph $L(G)$. The {\sc Efficient Edge Domination} (EED) problem asks for the existence of an e.e.d.s.\ in a given graph $G$. Thus, the EED problem for a graph $G$ corresponds to the ED problem for its line graph $L(G)$. Note that not every graph has an e.e.d.s. An efficient edge dominating set is also called \emph{dominating induced matching} ({\em d.i.m.} for short), and the EED problem is called the {\sc Dominating Induced Matching} (DIM) problem in various papers (see e.g. \cite{BraHunNev2010,BraMos2014,CarKorLoz2011,HerLozRieZamdeW2015,KorLozPur2014}); subsequently, we will use this notation instead of EED. 

In \cite{GriSlaSheHol1993}, it was shown that the DIM problem is \NP-complete; see also~\cite{BraHunNev2010,CarKorLoz2011,LuKoTan2002,LuTan1998}.
However, for various graph classes, DIM is solvable in polynomial time. For mentioning some examples, we need the following notions: 

Let $P_k$ denote the chordless path $P$ with $k$ vertices, say $a_1,\ldots,a_k$, and $k-1$ edges $a_ia_{i+1}$, $1 \le i \le k-1$; we also denote it as $P=(a_1,\ldots,a_k)$. 
 
For indices $i,j,k \ge 0$, let $S_{i,j,k}$ denote the graph $H$ with vertices $u,x_1,\ldots,x_i$, $y_1,\ldots,y_j$, $z_1,\ldots,z_k$ such that the subgraph induced by $u,x_1,\ldots,x_i$ forms a $P_{i+1}$ $(u,x_1,\ldots,x_i)$, the subgraph induced by $u,y_1,\ldots,y_j$ forms a $P_{j+1}$ $(u,y_1,\ldots,y_j)$, and the subgraph induced by $u,z_1,\ldots,z_k$ forms a $P_{k+1}$ $(u,z_1,\ldots,z_k)$, and there are no other edges in $S_{i,j,k}$; $u$ is called the {\em center} of $H$. 
Thus, {\em claw} is $S_{1,1,1}$, and $P_k$ is isomorphic to e.g.\ $S_{k-1,0,0}$.

For a set ${\cal F}$ of graphs, a graph $G$ is called {\em ${\cal F}$-free} if no induced subgraph of $G$ is contained in ${\cal F}$. 
If $|{\cal F}|=1$, say ${\cal F}=\{H\}$, then instead of $\{H\}$-free, $G$ is called $H$-free. 

\medskip

The following results are known: 

\begin{theorem}\label{DIMpolresults}
DIM is solvable in polynomial time for
\begin{itemize}
\item[$(i)$]  $S_{1,1,1}$-free graphs $\cite{CarKorLoz2011}$, 
\item[$(ii)$] $S_{1,2,3}$-free graphs $\cite{KorLozPur2014}$, 
\item[$(iii)$] $S_{2,2,2}$-free graphs $\cite{HerLozRieZamdeW2015}$,
\item[$(iv)$] $S_{2,2,3}$-free graphs $\cite{BraMos2017/2}$,
\item[$(v)$]  $P_7$-free graphs $\cite{BraMos2014}$ (in this case even in linear time), 
\item[$(vi)$] $P_8$-free graphs $\cite{BraMos2017}$.
\end{itemize}
\end{theorem}

In \cite{HerLozRieZamdeW2015}, it is conjectured that for every fixed $i,j,k$, DIM is solvable in polynomial time for $S_{i,j,k}$-free graphs (actually, an even stronger conjecture is mentioned in \cite{HerLozRieZamdeW2015}); this includes $P_k$-free graphs for $k \ge 8$.

\medskip

Based on the two distinct approaches described in \cite{BraMos2017} and in \cite{HerLozRieZamdeW2015,KorLozPur2014}, we show in this paper that DIM can be solved in polynomial time for $S_{1,2,4}$-free graphs (generalizing the corresponding results for $S_{1,2,3}$-free as well as for $P_7$-free graphs).

\section{Definitions and Basic Properties}\label{sec:basicnotionsresults}

\subsection{Basic notions}\label{subsec:basicnotions}

Let $G$ be a finite undirected graph without loops and multiple edges. Let $V(G)$ or $V$ denote its vertex set and $E(G)$ or $E$ its edge set; let $|V|=n$ and $|E|=m$.
For $v \in V$, let $N(v):=\{u \in V: uv \in E\}$ denote the {\em open neighborhood of $v$}, and let $N[v]:=N(v) \cup \{v\}$ denote the {\em closed neighborhood of $v$}. If $xy \in E$, we also say that $x$ and $y$ {\em see each other}, and if $xy \not\in E$, we say that $x$ and $y$ {\em miss each other}. A vertex set $S$ is {\em independent} in $G$ if for every pair of vertices $x,y \in S$, $xy \not\in E$. A vertex set $Q$ is a {\em clique} in $G$ if for every pair of vertices $x,y \in Q$, $x \neq y$, $xy \in E$. For $uv \in E$ let $N(uv):= N(u) \cup N(v) \setminus \{u,v\}$ and $N[uv]:= N[u] \cup N[v]$.

For $U \subseteq V$, let $G[U]$ denote the subgraph of $G$ induced by vertex set $U$. Clearly $xy \in E$ is an edge in $G[U]$ exactly when $x \in U$ and $y \in U$; thus, $G[U]$ can simply be denoted by $U$ (if understandable).

For $A \subseteq V$ and $B \subseteq V$, $A \cap B = \emptyset$, we say that $A \cojoin B$ ($A$ and $B$ {\em miss each other}) if there is no edge between $A$ and $B$, and $A$ and $B$ {\em see each other} if there is at least one edge between $A$ and $B$. If a vertex $u \notin B$ has a neighbor $v \in B$ then {\em $u$ contacts $B$}. If every vertex in $A$ sees every vertex in $B$, we denote it by $A \join B$. For $A=\{a\}$, we simply denote  $A \join B$ by $a \join B$, and correspondingly $A \cojoin B$ by $a \cojoin B$.
If for $A' \subseteq A$, $A' \cojoin (A \setminus A')$, we say that $A'$ is {\em isolated} in $G[A]$.
For graphs $H_1$, $H_2$ with disjoint vertex sets, $H_1+H_2$ denotes the disjoint union of $H_1$, $H_2$, and for $k \ge 2$, $kH$ denotes the disjoint union of $k$ copies of $H$. For example, $2P_2$ is the disjoint union of two edges. 

As already mentioned, a {\em chordless path} $P_k$, $k \ge 2$, has $k$ vertices, say $v_1,\ldots,v_k$, and $k-1$ edges $v_iv_{i+1}$, $1 \le i \le k-1$; 
the {\em length of $P_k$} is $k-1$. We also denote it as $P=(v_1,\ldots,v_k)$. 

A {\em chordless cycle} $C_k$, $k \ge 3$, has $k$ vertices, say $v_1,\ldots,v_k$, and $k$ edges $v_iv_{i+1}$, $1 \le i \le k-1$, and $v_kv_1$; the {\em length of $C_k$} is $k$.
  
Let $K_i$, $i \ge 1$, denote the clique with $i$ vertices. Let $K_4-e$ or {\em diamond} be the graph with four vertices, say $v_1,v_2,v_3,u$, such that $(v_1,v_2,v_3)$ forms a $P_3$ and $u \join \{v_1,v_2,v_3\}$; its {\em mid-edge} is the edge $uv_2$. 
A {\em gem} has five vertices say, $v_1,v_2,v_3,v_4,u$, such that $(v_1,v_2,v_3,v_4)$ forms a $P_4$ and $u \join \{v_1,v_2,v_3,v_4\}$.

A {\em butterfly} has five vertices, say, $v_1,v_2,v_3,v_4,u$, such that $v_1,v_2,v_3,v_4$ induce a $2P_2$ with edges $v_1v_2$ and $v_3v_4$ (the {\em peripheral edges} of the butterfly), and $u \join \{v_1,v_2,v_3,v_4\}$. 

We often consider an edge $e = uv$ to be a set of two vertices; then it makes sense to say, for example, $u \in e$ and $e \cap e' \neq \emptyset$, for an edge $e'$. For two vertices $x,y \in V$, let $dist_G(x,y)$ denote the {\em distance between $x$ and $y$ in $G$}, i.e., the length of a shortest path between $x$ and $y$ in $G$. 
The {\em distance between a vertex $z$ and an edge $xy$} is the length of a shortest path between $z$ and $x,y$, i.e., $dist_G(z,xy)= \min\{dist_G(z,v): v \in \{x,y\}\}$.
The {\em distance between two edges} $e,e' \in E$ is the length of a shortest path between $e$ and $e'$, i.e., $dist_G(e,e')= \min\{dist_G(u,v): u \in e, v \in e'\}$.
In particular, this means that $dist_G(e,e')=0$ if and only if $e \cap e' \neq \emptyset$. 

An edge subset $M \subseteq E$ is an {\em induced matching} if the pairwise distance between its members is at least 2, that is, $M$ is isomorphic to $kP_2$ for $k=|M|$. Obviously, if $M$ is a d.i.m.\ then $M$ is an induced matching.

Clearly, $G$ has a d.i.m.\ if and only if every connected component of $G$ has a d.i.m.; from now on, connected components are mentioned as {\em components}.

\subsection{Forbidden subgraphs and forced edges}\label{forbidsubgrforcededges}

The subsequent observations are helpful (some of them are mentioned e.g.\ in \cite{BraHunNev2010,BraMos2014,BraMos2017}). 

\begin{observation}[\cite{BraHunNev2010,BraMos2014}]\label{dimC3C5C7C4}
Let $M$ be a d.i.m.\ in $G$.
\begin{itemize}
\item[$(i)$] $M$ contains at least one edge of every odd cycle $C_{2k+1}$ in $G$, $k \ge 1$,
and exactly one edge of every odd cycle $C_3$, $C_5$, $C_7$ of $G$.
\item[$(ii)$] No edge of any $C_4$ can be in $M$.
\item[$(iii)$] For each $C_6$ either exactly two or none of its edges are in $M$.
\end{itemize}
\end{observation}

\noindent
{\bf Proof.} See e.g.\ Observation 2 in \cite{BraMos2014}.

\medskip

Since every triangle contains exactly one $M$-edge and no $M$-edge is in any $C_4$, and the pairwise distance of edges in any d.i.m.\ is at least 2, we obtain:

\begin{corollary}\label{cly:k4gemfree}
If a graph has a d.i.m.\ then it is $K_4$-free, gem-free and $\overline{C_k}$-free for any $k \ge 6$.
\end{corollary}

If an edge $e \in E$ is contained in {\bf every} d.i.m.\ of $G$, we call it a {\em forced} edge of $G$. If an edge $e \in E$ is {\bf not} contained in {\bf any} d.i.m.\ of $G$, we call it an {\em excluded} edge of $G$ (we can denote this by weight $w(e)=\infty$ or by coloring $e$ red). As a consequence of Observation~\ref{dimC3C5C7C4} $(ii)$, all edges in any $C_4$ of $G$ are excluded. Moreover, by Observation \ref{dimC3C5C7C4} $(i)$ for $C_3$, if for an edge $uv$ and a triangle $T$, $u \in V(T)$ and $v \notin V(T)$ then $uv$ is excluded. As another consequence of Observation \ref{dimC3C5C7C4} $(i)$ for $C_3$, we have:

\begin{observation}\label{obs:diamondbutterfly}
The mid-edge of any diamond in $G$ and the two peripheral edges of any induced butterfly are forced edges of $G$.
\end{observation}

Note that in a graph with d.i.m., the set of forced edges is an induced matching. Thus, our algorithm solving the DIM problem on $S_{1,2,4}$-free graphs has to check whether the set of forced edges is an induced matching. 

If $M$ is an induced matching of already collected forced edges and edge $vw$ is a new forced edge, we can reduce the graph as follows: 

\medskip

\noindent
{\bf Reduction-Step-($vw,M$).}
If $M \cup \{vw\}$ is not an induced matching then STOP - $G$ has no d.i.m., otherwise add $vw$ to $M$, i.e., $M:=M \cup \{vw\}$, delete $v$ and $w$ and all edges incident to $v$ and $w$ in $G$, and denote all edges that were at distance 1 from $vw$ in $G$ as excluded edges. 

\medskip

Obviously, the graph resulting from the reduction step is an induced subgraph of $G$. Recall that excluded edges are not in any d.i.m.\ of $G$.

\begin{observation}[\cite{BraMos2014}]\label{obs:redstep}
Let $M'$ be an induced matching which is a set of forced edges in $G$. Then $G$ has a d.i.m.\ $M$ if and only if after applying the reduction step to all edges in $M'$, the resulting graph has a d.i.m.\ $M \setminus M'$.
\end{observation}

Subsequently, this approach will often be used. Note that after applying the Reduction Step to all mid-edges of diamonds and all peripheral edges of butterflies in $G$, the resulting graph is (diamond, butterfly)-free. Moreover, by Corollary \ref{cly:k4gemfree}, a graph $G$ having a d.i.m.\ is $K_4$-free. Thus, from now on, we can assume that $G$ is connected $(K_4$, diamond, butterfly)-free.

Note that if $G$ has a d.i.m.\ $M$, and $V(M)$ denotes the vertex set of $M$ then $V \setminus V(M)$ is an independent set, say $I$, i.e., 
\begin{equation}\label{IV(M)partition}
V \mbox{ has the partition } V = I \cup V(M). 
\end{equation}

From now on, all vertices in $I$ are colored white and all vertices in $V(M)$ are colored black. According to \cite{HerLozRieZamdeW2015}, we also use the following notions: A partial black-white coloring of $V(G)$ is {\em feasible} if the set of white vertices is an independent set in $G$ and every black vertex has at most one black neighbor. A complete black-white coloring of $V(G)$ is {\em feasible} if the set of white vertices is an independent set in $G$ and every black vertex has exactly one black neighbor. Clearly, $M$ is a d.i.m.\ of $G$ if and only if the black vertices $V(M)$ and the white vertices $V \setminus V(M)$ form a complete feasible coloring of $V(G)$.  

\medskip

The Reduction Step mentioned above leads to a coloring reduction (i.e., C-reduction):

\medskip

\noindent
{\bf Edge C-Reduction.} Let $uw \in E(G)$. If $u$ and $w$ are black then 
\begin{itemize}
\item[$(i)$] color white all neighbors of $u$ and of $w$, and
\item[$(ii)$] remove $u$ and $w$ (and the edges containing $u$ or $w$) from $G$. 
\end{itemize}

Moreover, we have:

\medskip

\noindent
{\bf Vertex C-Reduction.} Let $u \in V(G)$. If $u$ is white, then
\begin{itemize}
\item[$(i)$] color black all neighbors of $u$, and
\item[$(ii)$] remove $u$ from $G$.
\end{itemize}

\subsection{The distance levels of an $M$-edge $xy$ in a $P_3$}\label{subsec:distlevels}

Based on \cite{BraMos2017}, we first describe some general structure properties for the distance levels of an edge in a d.i.m.\ $M$ of $G$.
Since $G$ is $(K_4$, diamond, butterfly)-free, we have:

\begin{observation}\label{obse:neighborhood}
For every vertex $v$ of $G$, $N(v)$ is the disjoint union of isolated vertices and at most one edge. Moreover, for every edge $xy \in E$, there is at most one common neighbor of $x$ and $y$.
\end{observation}

Since it is trivial to check whether $G$ has a d.i.m.\ $M$ with exactly one edge, from now on we can assume that $|M| \geq 2$. Since $G$ is connected and butterfly-free, we have:

\begin{observation}\label{obse:xy-in-P3}
If $|M| \geq 2$ then there is an edge in $M$ which is contained in a $P_3$ of $G$.
\end{observation}

Let $xy \in M$ be an $M$-edge for which there is a vertex $r$ such that $\{r,x,y\}$ induce a $P_3$ with edge $rx \in E$. This also means that $x$ and $y$ are black and lead to a feasible $xy$-coloring if there is indeed a d.i.m.\ $M$ of $G$ with $xy \in M$.
Let $N_0(xy):=\{x,y\}$ and for $i \ge 1$, let 
$$N_i(xy):=\{z \in V: dist_G(z,xy) = i\}$$
denote the {\em distance levels of $xy$}.
We consider a partition of $V$ into $N_i=N_i(xy)$, $i \ge 0$, with respect to the edge $xy$ (under the assumption that $xy \in M$).

Recall that by (\ref{IV(M)partition}), $V=I \cup V(M)$ is a partition of $V$ where $I$ is an independent set. Since we assume that $xy \in M$ (and is an edge in a $P_3$), clearly, $N_1 \subseteq I$ and thus:
\begin{equation}\label{N1subI}
N_1 \mbox{ is an independent set of white vertices.}
\end{equation}

Moreover, no edge between $N_1$ and $N_2$ is in $M$. Since $N_1 \subseteq I$ and all neighbors of vertices in $I$ are in $V(M)$, we have:
\begin{equation}\label{N2M2S2}
G[N_2] \mbox{ is the disjoint union of edges and isolated vertices. }
\end{equation}

Let $M_2$ denote the set of edges $uv \in E$ with $u,v \in N_2$ and let $S_2 = \{u_1,\ldots,u_k\}$ denote the set of isolated vertices in $N_2$; $N_2=V(M_2) \cup S_2$ is a partition of $N_2$. Obviously:
\begin{equation}\label{M2subM}
M_2 \subseteq M \mbox{ and } S_2 \subseteq V(M).
\end{equation}

If for $xy \in M$, an edge $e \in E$ is contained in {\bf every} dominating induced matching $M'$ of $G$ with $xy \in M'$, we say that $e$ is an {\em $xy$-forced} $M$-edge. The Reduction Step for forced edges can also be applied for $xy$-forced $M$-edges
(then, in the unsuccessful case, $G$ has no d.i.m.\ containing $xy$). Obviously, by (\ref{M2subM}), we have:
\begin{equation}\label{M2xymandatory}
\mbox{Every edge in } M_2 \mbox{ is an $xy$-forced $M$-edge}.
\end{equation}

Thus, from now on, after applying the Reduction Step for $M_2$-edges, we can assume that $M_2=\emptyset$, i.e., $N_2=S_2 = \{u_1,\ldots,u_k\}$. For every $i \in \{1,\ldots,k\}$, let $u'_i \in N_3$ denote the {\em $M$-mate} of $u_i$ (i.e., $u_iu'_i \in M$). Let $M_3=\{u_iu'_i: i \in \{1,\ldots,k\}\}$ denote the set of $M$-edges with one endpoint in $S_2$ (and the other endpoint in $N_3$). Obviously, by (\ref{M2subM}) and the distance condition for a d.i.m.\ $M$, the following holds:
\begin{equation}\label{noMedgesN3N4}
\mbox{ No edge with both ends in } N_3 \mbox{ and no edge between } N_3 \mbox{ and } N_4 \mbox{ is in } M.
\end{equation}

As a consequence of (\ref{noMedgesN3N4}) and the fact that every triangle contains exactly one $M$-edge (see Observation~\ref{dimC3C5C7C4} $(i)$), we have:
\begin{equation}\label{triangleaN3bcN4}
\mbox{For every triangle $abc$} \mbox{ with } a \in N_3, \mbox{ and } b,c \in N_4, \mbox{ $bc \in M$ is an $xy$-forced $M$-edge}.
\end{equation}

This means that for the edge $bc$, the Reduction Step can be applied, and from now on, we can assume that there is no such triangle $abc$ with $a \in N_3$ and $b,c \in N_4$, i.e., for every edge $uv \in E$ in $N_4$:
\begin{equation}\label{edgeN4N3neighb}
N(u) \cap N(v) \cap N_3 = \emptyset.
\end{equation}

\medskip

According to $(\ref{M2subM})$ and the assumption that $M_2=\emptyset$ (recall $N_2 = \{u_1,\ldots,u_k\}$), let:
\begin{enumerate}
\item[ ] $T_{one} := \{t \in N_3: |N(t) \cap N_2| = 1\}$;

\item[ ] $T_i := T_{one} \cap N(u_i)$, $i \in \{1,\ldots,k\}$;

\item[ ] $S_3 := N_3 \setminus T_{one}$.
\end{enumerate}

By definition, $T_i$ is the set of {\em private} neighbors of $u_i \in N_2$ in $N_3$ (note that $u'_i \in T_i$),
$T_1 \cup \ldots \cup T_k$ is a partition of $T_{one}$, and $T_{one} \cup S_3$ is a partition of~$N_3$.

\begin{lemma}[\cite{BraMos2017}]\label{lemm:structure2}
The following statements hold:
\begin{enumerate}
\item[$(i)$] For all $i \in \{1,\ldots,k\}$, $T_i \cap V(M)=\{u_i'\}$.
\item[$(ii)$] For all $i \in \{1,\ldots,k\}$, $T_i$ is the disjoint union of vertices and at most one edge.
\item[$(iii)$] $G[N_3]$ is bipartite.
\item[$(iv)$] $S_3 \subseteq I$, i.e., $S_3$ is an independent vertex set of white vertices.
\item[$(v)$] If a vertex $t_i \in T_i$ sees two vertices in $T_j$, $i \neq j$, $i,j \in \{1,\ldots,k\}$, then $u_it_i \in M$ is an $xy$-forced $M$-edge.
\end{enumerate}
\end{lemma}

\noindent
{\bf Proof.} $(i)$: Holds by definition of $T_i$ and by the distance condition of a d.i.m.\ $M$.

\noindent
$(ii)$: Holds by Observation \ref{obse:neighborhood}.

\noindent
$(iii)$: Follows by Observation \ref{dimC3C5C7C4} $(i)$ since every odd cycle in $G$ must contain at least one $M$-edge, and by (\ref{noMedgesN3N4}).

\noindent
$(iv)$: If $v \in S_3:= N_3 \setminus T_{one}$, i.e., $v$ sees at least two $M$-vertices then clearly, $v \in I$, and thus, $S_3 \subseteq I$ is an independent vertex set (recall that $I$ is an independent vertex set).

\noindent
$(v)$: Suppose that $t_1 \in T_1$ sees $a$ and $b$ in $T_2$. If $ab \in E$ then $u_2,a,b,t_1$ would induce a diamond in $G$. Thus, $ab \notin E$ and now,
$u_2,a,b,t_1$ induce a $C_4$ in $G$; by (\ref{noMedgesN3N4}), the only possible $M$-edge for dominating $t_1a,t_1b$ is $u_1t_1$, i.e., $t_1=u'_1$.
\qed

\medskip

Thus, by Lemma \ref{lemm:structure2} $(v)$, from now on, we can assume that for every $i,j \in \{1,\ldots,k\}$, $i \neq j$, any vertex $t_i \in T_i$ sees at most one vertex in $T_j$. In particular, if for some $i \in \{1,\ldots,k\}$, $T_i=\emptyset$ then there is no d.i.m.\ $M$ of $G$ with $xy \in M$. Thus, for every $i \in \{1,\ldots,k\}$, $T_i \neq \emptyset$.  

\begin{lemma}[\cite{BraMos2017}]\label{lemm:vwedgeS3}
The following statements hold:
\begin{enumerate}
\item[$(i)$] For every edge $vw \in E$ with $v,w \in N_3$, $vu_i \in E$, and $wu_j \in E$ $($possibly $i=j)$, we have $|\{v,w\} \cap \{u'_i,u'_j\}| = 1$.
\item[$(ii)$] For every edge $st \in E$ with $s \in S_3$ and $t \in T_i$, $t=u'_i$ holds, and thus $u_it$ is an $xy$-forced $M$-edge.
\end{enumerate}
\end{lemma}

\noindent
{\bf Proof.} 
$(i)$: By (\ref{noMedgesN3N4}), $N_3$ does not contain any $M$-edge, and clearly, if $vw \in E$ then either $v$ or $w$ is black; without loss of generality, let $v$ be black but then $v=u'_i$ and $w$ is white, i.e., $w \neq u'_j$.   

\noindent
$(ii)$: By Lemma \ref{lemm:structure2} $(iv)$, $S_3 \subseteq I$ and thus, by Lemma \ref{lemm:vwedgeS3} $(i)$, for the edge $st$ with $s \in S_3$, $s$ is white and thus, $t=u'_i$ holds.
\qed

\medskip

Subsequently, for checking if $G$ has a d.i.m.\ $M$ with $xy \in M$, we consider the cases $N_4 = \emptyset$ and $N_4 \neq \emptyset$.
In particular, we have the following property:

\begin{lemma}\label{endpointP5}
If $v \in N_i$ for $i \ge 3$ then $v$ is endpoint of a $P_5$, say with vertices $v,v_1,v_2,v_3,v_4$ such that $v_1,v_2,v_3,v_4 \in \{x,y\} \cup N_1 \cup \ldots \cup N_{i-1}$ and with edges $vv_1 \in E$, $v_1v_2 \in E$, $v_2v_3 \in E$, $v_3v_4 \in E$.
\end{lemma}

\noindent
{\bf Proof.} 
First assume that $v \in N_3$. Then $v$ has a neighbor $v_1 \in N_2$, and $v_1$ has a neighbor $v_2 \in N_1$. Since $xy$ is part of a $P_3$ with vertices $x,y,r$ and edges $xy, xr$, we have the following cases:

\begin{enumerate}
\item[(i)] $v_2=r$. Then for $x=v_3,y=v_4$, $v$ is endpoint of a $P_5$.
\item[(ii)] $v_2 \neq r$ and moreover, $v_1r \notin E$. If $v_2x \in E$ then, since $v_2r \notin E$ ($N_1$ is independent), we have a $P_5$ with endpoint $v$ and $v_3=x$, $v_4=r$, and if $v_2x \notin E$ but $v_2y \in E$, we again have a $P_5$ with endpoint $v$, and $v_3=y,v_4=x$.  
\end{enumerate}

If $v \in N_i$ for $i > 3$ then, if $i=4$, by similar arguments as above, and if $i > 4$, then obviously, $v$ is endpoint of a $P_5$ as claimed in the lemma. Thus, Lemma \ref{endpointP5} is shown.
\qed

\medskip

Let $X := \{x,y\} \cup N_1 \cup N_2 \cup N_3$ and $Y := V \setminus X$. 
Subsequently, for checking if $G$ has a d.i.m.\ $M$ with $xy \in M$, we first consider the possible colorings for $G[X]$.

\section{Coloring $G[X]$}\label{ColoringG[X]}

Recall that for every edge $uv \in M$, $u$ and $v$ are black, for $I=V(G) \setminus V(M)$, every vertex in $I$ is white, $N_2 = \{u_1,\ldots,u_k\}$ and all $u_i$, $1 \le i \le k$, are black, $T_i=N(u_i) \cap N_3$, and 

By Lemma \ref{lemm:structure2} $(iv)$ and the Vertex C-Reduction, we can assume that $S_3 = \emptyset$, i.e., $N_3=T_1 \cup \ldots \cup T_k$. Thus, no vertex in $N_3$ has two neighbors in $N_2$. 

Since no edge in $N_3$ is in $M$ (recall (\ref{noMedgesN3N4})), we have:
\begin{itemize}
\item[(R1)] All $N_3$-neighbors of a black vertex in $N_3$ must be colored white, and all $N_3$-neighbors of a white vertex in 
$N_3$ must be colored black.
\end{itemize}

Moreover, we have: 
\begin{itemize}
\item[(R2)] Every $T_i$, $i \in \{1,\ldots,k\}$, should contain exactly one vertex which is black. Thus, if $t_i \in T_i$ is black then all the remaining vertices of $T_i$ must be colored white.

\item[(R3)] If all but one vertices of $T_i$, $i \in \{1,\ldots,k\}$, are white and the final vertex $t$ is not yet colored, then $t$ must be colored black. 
\end{itemize}

Since no edge between $N_3$ and $N_4$ is in $M$ (recall (\ref{noMedgesN3N4})), we have:
\begin{itemize}
\item[(R4)] For every edge $st \in E$ with $t \in N_3$ and $s \in N_4$, $s$ is white if and only if $t$ is black and vice versa. 
\end{itemize}

Let us say that a vertex $t \in T_i$ (for $i \in \{1,\ldots,k\}$) is an {\em $N_3$-out-vertex} of $T_i$ if it is adjacent to some vertex of $T_j$ with $j \neq i$, 
$t$ is an {\em $N_4$-out-vertex} of $T_i$ if it is adjacent to some vertex of $N_4$, and is an {\em in-vertex} of $T_i$ otherwise. 
For finding a d.i.m.\ $M$ with $xy \in M$, one can remove all but one in-vertices (except for one of minimum weight); that can be done in polynomial time.
Thus, let us assume:

\begin{itemize}
\item[(A1)] For every $i \in \{1,\ldots,k\}$, $T_i$ has at most one in-vertex. 
\end{itemize}

Moreover, since no edge in a $C_4$ is in $M$ (recall Observation \ref{dimC3C5C7C4} $(ii)$) and since $u_i$ is black, we have:
\begin{itemize}
\item[(A2)] Each vertex of $T_i$ which belongs to an induced $C_4$ together with $u_i$ is colored by white; that can be done in polynomial time.
\end{itemize}

If  $|T_i|=1$, i.e., $T_i=\{t_i\}$, then $t_i$ is forced to be black, and $u_it_i$ is an $xy$-forced $M$-edge. Thus, after the Edge C-Reduction step (which again can be done in polynomial time), we can assume: 
\begin{itemize}
\item[(A3)] For every $i \in \{1,\ldots,k\}$, $|T_i| \geq 2$.  
\end{itemize}

\begin{lemma}\label{S124frcompN3N4outvertices}
If $T_i$ is already completely colored and if there is an edge between $T_i$ and $T_j$, $i \neq j$, then the color of all $N_3$-out-vertices as well as of all $N_4$-out-vertices of $T_j$ is forced by rules $(R1)-(R4)$.
\end{lemma}

\noindent
{\bf Proof.} Recall that by $(A3)$, $|T_i| \geq 2$ and $|T_j| \geq 2$. Let $bc \in E$ with $b \in T_i$ and $c \in T_j$. If $b$ is white then $c$ is black and thus, $T_j$ is completely colored. Now we assume that $b$ is black which implies that $c$ is white. Let $a \in T_i$ with $a \neq b$. Then $a$ is white and thus, $ac \notin E$. 
If there is no other $N_3$-out-vertex of $T_j$ then the in-vertex of $T_i$ is colored black, and thus, $T_j$ is completely colored.  

Now let $s \in T_j$, $s \neq c$, be another $N_3$-out-vertex of $T_j$, and assume that $s$ is not yet colored. Then, since all vertices of $T_i \cup \{c\}$ are already colored, we have $s \cojoin T_i \cup \{c\}$.
Then $s$ contacts some $T_h$ for $h \in \{1,\ldots,k\} \setminus \{i,j\}$, say $st \in E$ with $t \in T_h$. 

Again if $t$ contacts $T_i \cup \{c\}$, then $t$ (and thus $s$) is forced to have a color by (R1). Thus assume that $t \cojoin T_i \cup \{c\}$. 

If $T_j \setminus \{c,s\}$ contains only vertices which contact $T_i \cup \{c\}$ or which are adjacent to $t$, then the colors of all these vertices are forced either by (R1) or by (A2) (recalling that $G$ is diamond-free), and then the color of $s$ is forced by (R2) or (R3). Thus assume that there is a vertex $d \in T_j$ which does not contact $T_i \cup \{c,t\}$.

\medskip

Let $q_j \in N_1$ be a neighbor of $u_j$, and without loss of generality, assume that $q_jx \in E$. 

\medskip

First assume that $ab \notin E$. Then, since $u_j,d,s,t,c,b,u_i,a$ (with center $u_j$) do not induce an $S_{1,2,4}$, we have $ds \in E$.  

Since $u_j,q_j,s,t,c,b,u_i,a$ (with center $u_j$) do not induce an $S_{1,2,4}$, we have $q_ju_i \in E$. 

Since $q_j,x,u_i,a,u_j,s,t,u_h$ (with center $q_j$) do not induce an $S_{1,2,4}$, we have $q_ju_h \in E$. 

But then $q_j,x,u_i,a,u_h,t,s,d$ (with center $q_j$) induce an $S_{1,2,4}$, which is a contradiction.

\medskip

Thus $ab \in E$. Since $q_j,x,u_h,t,u_j,c,b,a$ (with center $q_j$) do not induce an $S_{1,2,4}$, we have $q_ju_h \notin E$. 

Since $q_j,x,u_i,a,u_j,s,t,u_h$ (with center $q_j$) do not induce an $S_{1,2,4}$, we have $q_ju_i \notin E$; let $q_i \in N_1$ be a neighbor of $u_i$, and by the same argument, we have $q_iu_j \notin E$. 

But now, $u_j,q_j,s,t,c,b,u_i,q_i$ (with center $u_j$) induce an $S_{1,2,4}$, which is a contradiction. 

\medskip

Now let $s$ be any $N_4$-out-vertex of $T_j$ (for $s \neq c$), and again assume that $s$ is not yet colored. Then, since all vertices of $T_i \cup \{c\}$ are already colored, we have $s \cojoin T_i \cup \{c\}$. Let $z \in N_4$ be a neighbor of $s$.

If $z$ contacts $T_i \cup \{c\}$, then by (R4), the color of $z$ and the color of $s$ are forced. Then assume that $z$ does not contact $T_i \cup \{c\}$, i.e., 
$z \cojoin T_i \cup \{c\}$.

If $z$ has degree 1, then by Proposition \ref{isolatednonadjtotriangle}, the color of $s$ is forced to be black. Thus, we assume that the degree of $z$ is at least 2; let $z'$ be a new neighbor of $z$. 
Since $z \cojoin T_i \cup \{c\}$, we have $z' \notin T_i$, and we can assume that $z' \notin T_j$ (else $u_j,s,z,z'$ would induce a diamond - which is impossible - or $C_4$ which implies that $s$ is forced to be white). 

If $s,z,z'$ induce a triangle, then recall that by (\ref{noMedgesN3N4}), no edge between $N_3$ and $N_4$ as well as no edge in $N_3$ is in $M$, but every triangle contains exactly one $M$-edge; if $z' \in N_4$ and $s,z,z'$ induce a triangle then $zz'$ is an $xy$-forced $M$-edge.

Thus, assume that $z's \notin E$. 

\medskip

If there is a common neighbor $q \in N_1$ such that $qu_i \in E$ and $qu_j \in E$ (and without loss of generality, $qx \in E$), then, 
since $q,x,u_i,a,u_j,s,z,z'$ (with center $q$) do not induce an $S_{1,2,4}$, we have $z'a \in E$, and analogously, $z'b \in E$. 
Since $u_i,a,b,z'$ do not induce a diamond, we have $ab \notin E$, but now, $u_i,a,b,z'$ induce a $C_4$, which is a contradiction for the fact that $b$ is black. 

\medskip

Thus, $u_i$ and $u_j$ do not have a common neighbor in $N_1$; let $q_i \in N_1$ with $q_iu_i \in E$, and $q_j \in N_1$ with  $q_ju_j \in E$, $q_i \neq q_j$.
But then $u_j,q_j,s,z,c,b,u_i,q_i$ (with center $u_j$) induce an $S_{1,2,4}$, which is a contradiction. 

\medskip

Thus, Lemma \ref{S124frcompN3N4outvertices} is shown.
\qed

\subsection{The Case $N_4 = \emptyset$}\label{N4empty}

Recall that $S_3 = \emptyset$, i.e., $N_3=T_1 \cup \ldots \cup T_k$. 

\medskip

$G[\{u_i\} \cup T_i]$ is a {\em trivial component in $G[S_2 \cup N_3]$} if $T_i$ has no contact to any other $T_j$, $j \neq i$. By the way, if $T_i=\emptyset$, it leads to a contradiction. 
Obviously, checking a possible d.i.m.\ $M$ with $xy \in M$ can be done easily (and independently) for trivial components; for a minimum weight vertex $u'_i \in T_i$ let $u_iu'_i \in M$.  

\medskip

From now on we present a coloring procedure for nontrivial components $K$ in $G[S_2 \cup N_3]$, i.e., $K$ contains at least two $T_i,T_j$, $i \neq j$ with contact to each other. 

\medskip

For any nontrivial component $K$ in $G[S_2 \cup N_3]$, say $V(K)=\{u_1,\ldots,u_p\} \cup T_1 \cup \ldots \cup T_p$, $p \ge 2$, the coloring procedure starts with at most $|T_1|$ possible colorings of $T_1$ (recall (R2)). Then for each of these possible colorings, by Lemma \ref{S124frcompN3N4outvertices}, it can be applied to $T_i$ which contacts $T_1$ etc.\ until all vertices in $K$ are feasibly colored or it leads to a contradiction. Since by (A1), there is at most one in-vertex of $T_i$, the color of such an in-vertex is finally forced by (R2) and (R3) and by Lemma \ref{S124frcompN3N4outvertices}.

As an example of a contradiction, if there are three edges between $T_1$ and $T_2$, say $t_1t_2 \in E$, $t'_1t'_2 \in E$, and $t''_1t''_2 \in E$ for $t_i,t'_i,t''_i \in T_i$, $i=1,2$, then $t_1$ is black if and only if $t_2$ is white, $t'_1$ is black if and only if $t'_2$ is white, and 
$t''_1$ is black if and only if $t''_2$ is white. Without loss of generality, assume that $t_1$ is black, and $t_2$ is white. Then $t'_1$ is white, and $t'_2$ is black, but now, $t''_1$ and $t''_2$ are white which leads to a contradiction. 
Then, by the contradiction, $xy \notin M$ for any dominating induced matching $M$ of $G$.

\medskip

If the coloring procedure for $K$ ends without contradiction with respect to some of the $|T_1|$ possible 
colorings of $T_1$ then we choose a minimum weight solution for the DIM problem on $K$. 

\medskip

If for at least one of the components, it leads to a contradiction then there is no such d.i.m.\ $M$ with $xy \in M$. 

\begin{corollary}\label{S124frcompN4emptypolcol}
If $N_4 = \emptyset$ then the DIM problem can be done in polynomial time. 
\end{corollary}

\noindent
{\bf Proof.}
For trivial components, it can be obviously done. For every nontrivial component, it can be done in polynomial time as above. Thus,  
in the case $N_4 = \emptyset$, it leads to a polynomial time solution since all the components of $G[S_2 \cup N_3]$ can be independently colored. 
\qed

\subsection{The Case $N_4 \neq \emptyset$}\label{N4nonempty}

Recall again that $S_3=\emptyset$ and $N_3=T_1 \cup \ldots \cup T_k$.

\begin{proposition}\label{isolatednonadjtotriangle}
If $z \in N_4$ is isolated in $G[Y]$ and $z$ contacts $t_i \in T_i$ then $u_it_i \in M$ is an $xy$-forced $M$-edge. In particular, if $|N(z) \cap T_i| \ge 2$ then $G$ has no d.i.m.\ $M$ with $xy \in M$.   
\end{proposition}

\noindent
{\em Proof.} Clearly, if $xy \in M$, there is an $M$-edge $u_it'_i$ for some $t'_i \in T_i$. However, if $t'_i \neq t_i$ then, since $z$ is isolated, the only possible way of dominating edge $t_iz$ is $u_it_i \in M$. If $|N(z) \cap T_i| \ge 2$, say $a,b \in T_i$ with $az \in E$, $bz \in E$ then, since $G$ is diamond-free, $ab \notin E$ but then $u_i,a,b,z$ induce a $C_4$, and by Observation \ref{dimC3C5C7C4} $(ii)$, there is no $M$-edge dominating $az$, $bz$. Thus, in this case, $G$ has no d.i.m.\ $M$ with $xy \in M$.   
\qed

\medskip

Again, since by (A1), there is at most one in-vertex of $T_j$, the color of such an in-vertex is finally forced by (R2) and (R3) and by 
Lemma \ref{S124frcompN3N4outvertices}.

\begin{corollary}\label{S124frcompN4nonemptypolcol}
If $N_4 \neq \emptyset$ then again for every component in $G[S_2 \cup N_3]$, it can be done in polynomial time whether it could be feasibly colored or it could lead to a contradiction.
\end{corollary}

For combining the ``coloring approach'' of \cite{HerLozRieZamdeW2015,KorLozPur2014} with the above results, we show: 
\begin{lemma}\label{lemm:S124Xcoloring}
For $S_{1,2,4}$-free graphs $G$, the number of feasible $xy$-colorings of $G[X]$ is at most polynomial. In particular, such $xy$-colorings can be detected in polynomial time.
\end{lemma}

For the proof of Lemma \ref{lemm:S124Xcoloring}, we will collect some propositions below. 

\medskip

Connecting a feasible coloring of $G[X]$ with a corresponding one of $G[Y]$ means that every vertex $v \in N_3$ determines the color of its (possible) neighbors in $N_4$: Clearly, if $v \in N_3$ is white then all of its neighbors in $N_4$ are forced to be black, and by fact (\ref{noMedgesN3N4}), if $v \in N_3$ is black then all of its neighbors in $N_4$ are forced to be white. Clearly, it can result in a contradiction, e.g., if a vertex $u \in N_4$ is adjacent to a white vertex $w \in N_3$ and to a black vertex $v \in N_3$. Thus, in this case, $xy$ is not contained in any d.i.m.\ of $G$.

\medskip

Recall that we have a partial feasible $xy$-coloring which means that $x$ and $y$ are black, all vertices of $N_1$ are white, all vertices of $N_2= \{u_1,\ldots,u_k\}$ are black, and we can assume that $S_3=\emptyset$ (recall that by Lemma \ref{lemm:structure2}~$(iv)$, any vertex in the component $K$ contacting $S_3$ is black, and thus, the color of each vertex of $K$ is forced, and every vertex in $N_4$ contacting $S_3$ is black). 
Then let us see how this partial feasible $xy$-coloring can be extended.

\medskip

By Lemma \ref{lemm:structure2} we have:

\begin{proposition}\label{QcompK}
Let $Q$ denote the family of components of $G[S_2 \cup T_{one}]$, and let $K$ be a member of $Q$. 
\begin{itemize}
\item[$(i)$] If for some $i \in \{1,\ldots,k\}$, $K$ contains a subset $T_i$ such that $|T_i| \geq 2$ and there is a vertex $z \in N_4$ with $z \join T_i$ 
 then, by the $C_4$-property in Observation $\ref{dimC3C5C7C4}$ $(ii)$ and since $G$ is diamond-free, $G$ has no d.i.m.\ with $xy \in M$.

\item[$(ii)$] If $K \cojoin N_4$ then, by the results of Section $\ref{N4empty}$, $K$ can be treated independently to the other members of $Q$. 
\end{itemize}
\end{proposition}

\medskip

Thus, by the previous rules and assumptions as well as Propositions \ref{isolatednonadjtotriangle} and \ref{QcompK}, we can restrict $Q$ as follows: Let $Q^*$ be the family of components $H$ of $G[S_2 \cup T_{one}]$ such that: 
\begin{itemize}
\item[(R5)] for any $z \in N_4$, there is at least one non-neighbor of $z$ in $V(H) \cap N_3$, 
\item[(R6)] some vertex of $V(H)$ contacts $N_4$, and  
\item[(R7)] no vertex $z \in N_4$ is isolated in $G[Y]$.
\end{itemize}

\begin{proposition}\label{P1S124}
If a vertex of $N_4$ contacts at least two members of $Q^*$ then $|Q^*| \le 3$. 
\end{proposition}

\noindent
{\em Proof.} Suppose to the contrary that there are four distinct members $H_1,H_2,H_3,H_4$ of $Q^*$ such that a vertex $z \in N_4$ contacts $H_1$ and $H_2$. 
Let $u_i \in V(H_i) \cap S_2$, $1 \le i \le 4$, such that for $i \in \{1,2\}$, there are $t_i \in V(H_i) \cap T_i$ with $zt_i \in E$, and for $i \in \{3,4\}$, 
there are $t_i \in V(H_i) \cap T_i$ with $zt_i \notin E$ (such non-neighbors $t_3,t_4$ of $z$ exist by condition (R5) of the definition of $Q^*$). 

Clearly, for any $i \neq j$, we have $t_it_j \notin E$ since $t_1,\ldots,t_4$ are in distinct components $H_1,\ldots,H_4$, and clearly $t_iu_j \notin E$.

\begin{clai}\label{nocommonN1neighuiuj}
For all $i,j \in \{1,2,3,4\}$, $i \neq j$, $u_i$ and $u_j$ do not have any common neighbor in $N_1$.  
\end{clai}

\noindent
{\em Proof.} We first claim that $u_1$ and $u_3$ do not have a common neighbor in $N_1$: Let $a_1 \in N_1$ with $a_1u_1 \in E$, and without loss of generality, let $a_1x \in E$. Since $a_1,x,u_3,t_3,u_1,t_1,z,t_2$ (with center $a_1$) do not induce an $S_{1,2,4}$, we have $a_1u_3 \notin E$, and thus, $u_1$ and $u_3$ do not have any common neighbor in $N_1$.  

Similarly, by symmetry, we can show that $u_1$ and $u_4$ (respectively, $u_2$ and $u_3$, $u_2$ and $u_4$) do not have any common neighbor in $N_1$. 
$\diamond$

\medskip

Let $a_1 \in N_1$ be adjacent to $u_1$, and let $a_3 \in N_1$ be adjacent to $u_3$. By the previous facts, $a_3$ is nonadjacent to $u_1,u_2$ and $a_1$ is nonadjacent to $u_3,u_4$. 

Next we claim that $a_1$ is nonadjacent to $u_2$: Otherwise an $S_{1,2,4}$ arises of center $a_1$ with four vertices in $\{x,y,a_3,u_3,t_3\}$, and $u_1,t_1$, and $u_2$. Then by construction, let $a_2 \in N_1$ be adjacent to $u_2$. By the previous facts, $a_2$ is nonadjacent to $u_1,u_3,u_4$.

Now, $a_1$ and $a_2$ are nonadjacent to $u_4$. Furthermore $a_3$ is nonadjacent to $u_4$, since otherwise an $S_{1,2,4}$ arises of center $a_3$ with four vertices in $\{x,y,a_1,u_1,t_1\}$, and $u_3,t_3$, and $u_4$. By construction, let $a_4 \in N_1$ be adjacent to $u_4$; recall that $a_1,a_2,a_3,a_4$ are pairwise distinct and $a_4$ is nonadjacent to $u_1,u_2,u_3$. 
$\diamond$

\medskip

Since $G$ is diamond-free, at most one of $a_1,a_2,a_3,a_4$ is adjacent to $x$ and to $y$. Without loss of generality, assume $xa_1 \in E$. If $x$ has at least three neighbors in $a_1,a_2,a_3,a_4$, say additionally, $xa_i \in E$ and $xa_j \in E$, $i \neq j$ and $i,j \neq 1$, then $x,a_j,a_i,u_i,a_1,u_1,t_1,z$ (with center $x$) induce an $S_{1,2,4}$. Now assume that $x$ misses at least two of $a_2,a_3,a_4$. Analogously, if $y$ is adjacent to at least three of $a_1,a_2,a_3,a_4$ then we get an $S_{1,2,4}$ as above. Finally, if each of $x$ and $y$ has exactly two neighbors in $a_1,a_2,a_3,a_4$ (but no common neighbor since each of $a_i$ is adjacent to $x$ or $y$) then one can easily check that there is an $S_{1,2,4}$. Thus, Proposition~\ref{P1S124} is shown.
\qed

\medskip

Clearly, in the case $|Q^*| \le 3$, the number of $xy$-colorings of $G[X]$ is bounded by a polynomial.
From now on, by Proposition \ref{P1S124}, we can add another restriction:

\begin{itemize}
\item[(R8)] Each vertex of $N_4$ contacts at most one member of $Q^*$.
\end{itemize}

Let $Q^{**}$ be the family of components $H$ of $G[S_2 \cup T_{one}]$ fulfilling conditions (R5)--(R8).
 
\begin{proposition}\label{notisoQatmost3}
$|Q^{**}| \le 3$.
\end{proposition}

\noindent
{\em Proof.} Suppose to the contrary that there are four distinct members $H_1,H_2,H_3,H_4$ of $Q^{**}$ such that a non-isolated vertex $z_1 \in N_4$ contacts $H_1$. Let again $u_i \in V(H_i) \cap S_2$, $1 \le i \le 4$, let $z_1$ contact $T_1$, and since $z_1$ is not isolated in $G[Y]$, there is a neighbor $z_2 \in Y$ of $z_1$. By condition (R8), $z_1 \cojoin T_i$, $i \ge 2$, and by condition (R5), $z_2$ has a non-neighbor in each $T_i$, $i \ge 2$. Let $a_1 \in N_1$ be a neighbor of $u_1$, and without loss of generality, let $a_1x \in E$. 

 For $i \ge 2$, let $t_i \in T_i$ be a non-neighbor of $z_2$. Then, since $a_1,x,u_i,t_i,u_1,t_1,z_1,z_2$ (with center $a_1$) do not induce an $S_{1,2,4}$, we have $a_1u_i \notin E$ for each $i \ge 2$. Let $a_i \in N_1$ be a neighbor of $u_i$. Analogously, $a_iu_1 \notin E$ for each $i \ge 2$. If $u_i$ and $u_j$, $i \neq j$, have a common neighbor $a_i \in N_1$ then it is easy to see that there is an $S_{1,2,4}$. Thus we can assume that each $u_i$ has its private neighbor $a_i$ in $N_1$. Now, since $G$ is diamond-free, at most one of $a_i$ is adjacent to $x$ and to $y$, and thus, as in the proof of Proposition~\ref{P1S124}, it is easy to see that this again leads to an $S_{1,2,4}$. Thus, Proposition \ref{notisoQatmost3} is shown.
\qed

\medskip

\noindent
{\bf Proof of Lemma \ref{lemm:S124Xcoloring}.} It follows by Propositions \ref{isolatednonadjtotriangle}--\ref{notisoQatmost3}. 
In particular all the above properties can be checked in polynomial time. 
\qed

\section{The Structure of $G[Y]$}\label{StructureG[Y]}

Recall $X:=\{x,y\} \cup N_1 \cup N_2 \cup N_3$ and $Y:= V \setminus X$. Clearly, in this section, $Y \neq \emptyset$. 
We show that $G[Y]$ is $S_{1,2,2}$-free. 
In Section \ref{sec:DIMpolS124fr}, for coloring $G[Y]$, we will use the polynomial time result for DIM on $S_{1,2,2}$-free graphs (see Theorem \ref{DIMpolresults} $(ii)$). 
The approach in \cite{HerLozRieZamdeW2015,KorLozPur2014}, however, is strongly based on coloring vertices white or black as already mentioned (i.e., all vertices of $V(M)$ are black and all vertices of $I=V \setminus V(M)$ are white). 
By \cite{KorLozPur2014}, for $S_{1,2,3}$-free graphs, DIM is also solvable in polynomial time if $G$ has a special subset of vertices whose colors are fixed to be black or white.

\begin{lemma}\label{S124frYS122fr}
If $G$ is $S_{1,2,4}$-free then $G[Y]$ is $S_{1,2,2}$-free.
\end{lemma}

\noindent
{\bf Proof.} 
Suppose to the contrary that there is an $S_{1,2,2}$ $H$ in $G[Y]$, say with vertices $d,a_1,a_2,b_1,b_2,c_1$ and edges $da_1 \in E$, $db_1 \in E$, $dc_1 \in E$, $a_1a_2 \in E$, $b_1b_2 \in E$. Let $v \in N_p$ be a neighbor of $H$ with smallest $p \ge 3$ (such a neighbor exists since $G$ is connected). 
By Lemma \ref{endpointP5}, $v$ is endpoint of a $P_5$, say $P(v)$ with vertices $v,v_1,v_2,v_3,v_4$, and clearly, none of $v_i$, $i \in \{1,2,3,4\}$, is a neighbor of $H$.   

We first claim: 
\begin{equation}\label{vnonadjd}
vd \notin E. 
\end{equation}

\noindent    
{\em Proof.} Suppose to the contrary that $vd \in E$. First assume that $va_1 \in E$. Then, since $G$ is diamond-free, $va_2 \notin E$, $vb_1 \notin E$, and $vc_1 \notin E$. 

If $vb_2 \in E$ then $v,a_1,b_2,b_1,v_1,v_2,v_3,v_4$ (with center $v$) would induce an $S_{1,2,4}$. Thus, $vb_2 \notin E$ but now, 
$d,c_1,b_1,b_2,v,v_1,v_2,v_3$  (with center $d$) induce an $S_{1,2,4}$, which is a contradiction.
By symmetry, the same arguments hold if $vb_1 \in E$ (instead of $va_1 \in E$). 

\medskip
 
From now on, let $va_1 \notin E$ and $vb_1 \notin E$. If $va_2 \notin E$ then $d,b_1,a_1,a_2,v,v_1,v_2,v_3$ (with center $d$) would induce an $S_{1,2,4}$, and similarly if $vb_2 \notin E$. Thus, $va_2 \in E$ and $vb_2 \in E$ but now, $v,b_2,a_2,a_1,v_1,v_2,v_3,v_4$ (with center $v$) induce an $S_{1,2,4}$,  
which is a contradiction.

Thus, (\ref{vnonadjd}) is shown.
$\diamond$

\medskip

Next we claim: 
\begin{equation}\label{vnonadja1a2}
(va_1 \notin E \mbox{ or } va_2 \notin E) \mbox{ and } (vb_1 \notin E \mbox{ or } vb_2 \notin E). 
\end{equation}

\noindent
{\em Proof.} Suppose to the contrary that $va_1 \in E$ and $va_2 \in E$. Then, since $G$ is butterfly-free, $vb_1 \notin E$ or $vb_2 \notin E$.

If $vb_1 \in E$ then $vb_2 \notin E$, and thus, $v,a_1,b_1,b_2,v_1,v_2,v_3,v_4$ (with center $v$) would induce an $S_{1,2,4}$. 
Analogously, if $vb_2 \in E$ then $vb_1 \notin E$, and similarly, $v,a_1,b_2,b_1,v_1,v_2,v_3,v_4$ (with center $v$) would induce an $S_{1,2,4}$,  
which is a contradiction in each case.

Thus, $vb_1 \notin E$ and $vb_2 \notin E$. Now, since $d,c_1,b_1,b_2,a_1,v,v_1,v_2$ (with center $d$) does not induce an $S_{1,2,4}$, we have $vc_1 \in E$ but now, 
$v,a_2,c_1,d,v_1,v_2,v_3,v_4$ (with center $v$) induce an $S_{1,2,4}$, which is a contradiction. 

\medskip

By symmetry, also $vb_1 \in E$ and $vb_2 \in E$ is impossible. Thus, (\ref{vnonadja1a2}) is shown.
$\diamond$

\medskip

If $v$ has exactly one neighbor in $a_1,a_2$ and exactly one neighbor in $b_1,b_2$, say $va_1 \in E$ and $vb_1 \in E$, then, by (\ref{vnonadja1a2}), 
$v,b_1,a_1,a_2,v_1,v_2,v_3,v_4$ (with center $v$) would induce an $S_{1,2,4}$, and similarly in every other case. 
Thus, without loss of generality assume that $va_1 \notin E$ and $va_2 \notin E$. By (\ref{vnonadja1a2}), $v$ sees at most one of $b_1,b_2$.

If $vb_1 \in E$ (and $vb_2 \notin E$) then $b_1,b_2,d,a_1,v,v_1,v_2,v_3$ (with center $b_1$) would induce an $S_{1,2,4}$. 
If $vb_2 \in E$ (and $vb_1 \notin E$) and if $vc_1 \notin E$ then $d,c_1,a_1,a_2,b_1,b_2,v,v_1$ (with center $d$) would induce an $S_{1,2,4}$. 
Thus, $vc_1 \in E$, but now, $d,b_1,a_1,a_2,c_1,v,v_1,v_2$ (with center $d$) induce an $S_{1,2,4}$, which is a contradiction. 
Thus, $vb_1 \notin E$ and $vb_2 \notin E$. 

\medskip

Finally, $c_1$ is the only neighbor of $v$ in $H$ but then again $d,b_1,a_1,a_2,c_1,v,v_1,v_2$ (with center $d$) induce an $S_{1,2,4}$, which is a contradiction. Thus, Lemma \ref{S124frYS122fr} is shown.    
\qed

\medskip

For the case of $S_{1,1,4}$-free graphs, we can show even more:

\begin{lemma}\label{S114frYS111fr}
If $G$ is $S_{1,1,4}$-free then $G[Y]$ is $S_{1,1,1}$-free.
\end{lemma}

\noindent
{\bf Proof.} 
Suppose to the contrary that there is an $S_{1,1,1}$ $H$ in $G[Y]$, say with vertices $d,a,b,c$ and edges $da \in E$, $db \in E$, $dc \in E$.
Let $v \in N_p$ be a neighbor of $H$ with smallest $p \ge 3$ (such a neighbor exists since $G$ is connected). As above, by Lemma \ref{endpointP5}, $v$ is endpoint of a $P_5$, say $P(v)$ with vertices $v,v_1,v_2,v_3,v_4$, and clearly, none of $v_i$, $i \in \{1,2,3,4\}$, is a neighbor of~$H$.   
We first claim: 
\begin{equation}\label{vnonadjmidpointclaw}
vd \notin E. 
\end{equation}

\noindent   
{\em Proof.} Suppose to the contrary that $vd \in E$. If $va \in E$ then, since $G$ is diamond-free, $vb \notin E$, and $vc \notin E$, but now, 
 $d,b,c,v,v_1,v_2,v_3$ (with center $d$) would induce an $S_{1,1,4}$. 
By symmetry, the same arguments hold if $vb \in E$ or $vc \in E$. Thus, (\ref{vnonadjmidpointclaw}) is shown.
$\diamond$

\medskip

If $v$ is adjacent to only one of $a,b,c$, say $va \in E$, then $d,b,c,a,v,v_1,v_2$ (with center $d$) would induce an $S_{1,1,4}$. 
Thus, $v$ is adjacent to at least two of $a,b,c$, say $va \in E$ and $vb \in E$ but then $v,a,b,v_1,v_2,v_3,v_4$ (with center $v$) induce an $S_{1,1,4}$, which is a contradiction. Thus, Lemma \ref{S114frYS111fr} is shown. 
\qed 

\section{A polynomial-time algorithm for DIM on $S_{1,2,4}$-free graphs}\label{sec:DIMpolS124fr}

The following procedure is part of the algorithm:

\medskip

\begin{proc}[DIM-with-$xy$-in-$S_{1,2,4}$-free-graphs]\label{DIMwithxyS124}

\begin{tabbing}	   
xxxxxxx \= \kill\\
{\bf Input:} \> A connected $(S_{1,2,4},K_4$,diamond,butterfly$)$-free graph $G = (V,E)$, and\\ 
\> an edge $xy \in E$ which is part of a $P_3$ in $G$.\\
{\bf Task:} \> Return a d.i.m.\ $M$ with $xy \in M$ $($STOP with success$)$ or\\ 
\> a proof that $G$ has no d.i.m.\ $M$ with $xy \in M$ $($STOP with failure$)$. 
\end{tabbing}

\begin{itemize}

\item[$(a)$] Set $M:= \{xy\}$. Determine the distance levels $N_i = N_i(xy)$, $i \ge 1$, with respect to $xy$.

\item[$(b)$] Check whether $N_1$ is an independent set $($see fact $(\ref{N1subI}))$ and $G[N_2]$ is the disjoint union of edges and isolated vertices $($see fact $(\ref{N2M2S2}))$. If not, then STOP with failure.

\item[$(c)$] For the set $M_2$ of edges in $G[N_2]$, apply the Edge C-Reduction for every edge in $M_2$ correspondingly. Moreover, apply the Edge C-Reduction for each edge $bc$ according to fact $(\ref{triangleaN3bcN4})$ and then for each edge $u_it_i$ according to Lemma $\ref{lemm:structure2}$ $(v)$.

\item[$(d)$] {\bf if} $N_4 = \emptyset$ then apply the approach described in Section $\ref{N4empty}$. Then either return that $G$ has no d.i.m.\ $M$ with $xy \in M$ or return $M$ as a d.i.m.\ with $xy \in M$. 

\item[$(e)$] {\bf if} $N_4 \neq \emptyset$ {\bf then} for $X := \{x,y\} \cup N_1 \cup N_2 \cup N_3$ and $Y := V \setminus X$ {\bf do}: 

\begin{itemize}
\item[$(e.1)$] According to Lemma $\ref{lemm:S124Xcoloring}$, compute all feasible $xy$-colorings of $G[X]$. If no such $xy$-coloring exists, then STOP with failure. 

\item[$(e.2)$]  {\bf for each} feasible $xy$-coloring of $G[X]$  {\bf do}: 
\begin{enumerate}
\item[$(e.2.1)$]  Derive a partial coloring of $G[Y]$ by the forcing rules; {\bf if} a contradiction arises in vertex coloring {\bf then} STOP with failure. 

\item[$(e.2.2)$] According to Lemma $\ref{S124frYS122fr}$, apply a polynomial time algorithm for DIM on $S_{1,2,2}$-free graphs $($see $\cite{KorLozPur2014})$ for $G[Y]$ with its partial coloring; 
{\bf if} it returns a d.i.m.\ of $G[Y]$ {\bf then} STOP with success and return the feasible $xy$-coloring of $G$ derived by the feasible $xy$-coloring of $G[X]$ and by such a d.i.m.\ of $G[Y]$. 
\end{enumerate}

\item[$(e.3)$] STOP with failure.
\end{itemize}
\end{itemize}
\end{proc}

\begin{theorem}\label{theo:procedureDIMxyS124}
Procedure $\ref{DIMwithxyS124}$ is correct and can be done in polynomial time.
\end{theorem}

\noindent
{\bf Proof.} The correctness of the procedure follows from the structural analysis of $S_{1,2,4}$-free graphs with a d.i.m.

\medskip

The polynomial time bound follows from the fact that Steps (a) and (b) can clearly be done in polynomial time, Step (c) can be done in polynomial time since the Edge C-Reduction can be done in polynomial time, Steps (d) and (e) can be done in polynomial time by the results in Sections \ref{ColoringG[X]} and 
\ref{StructureG[Y]} and by the fact that DIM can be solved in polynomial time for $S_{1,2,2}$-free graphs \cite{KorLozPur2014} (see also \cite{HerLozRieZamdeW2015}).
\qed 

\begin{algo}[DIM-$S_{1,2,4}$-free]\label{DIMS124fr}

\begin{tabbing}	   
xxxxxxx \= \kill\\
{\bf Input:} \> A connected $(S_{1,2,4},K_4)$-free graph $G = (V,E)$. \\
{\bf Task:} \> Determine a d.i.m.\ of $G$ if there is one, or find out that $G$ has no d.i.m.
\end{tabbing}

\begin{itemize}
\item[$(A)$] Determine the set $F_1$ of all mid-edges of diamonds in $G$, and the set $F_2$ of all peripheral edges of butterflies in $G$. Let $M:=F_1 \cup F_2$. Check whether $M$ is an induced matching in $G$. If not then STOP--$G$ has no d.i.m. Otherwise, check whether $M$ is a dominating edge set of $G$. If yes, we are done. Otherwise apply the Edge C-Reduction for every edge in $F_1 \cup F_2$; without loss of generality, assume that the resulting graph $G'=(V',E')$ is connected (if not, do the next steps for each component of $G'$). Let $G:=G'$.

$\{$From now on, $G$ is $(S_{1,2,4},K_4$,diamond,butterfly)-free.$\}$

\item[$(B)$] Check whether $G$ has a single edge $uv \in E$ which is a d.i.m.\ of $G$. If yes then select such an edge as output and STOP--this is a d.i.m.\ of $G$. 
$\{$Otherwise, every d.i.m.\ of $G$ would have at least two edges.$\}$

\item[$(C)$] {\bf for each} edge $xy \in E$ in a $P_3$ of $G$, carry out Procedure $\ref{DIMwithxyS124}$; {\bf if} it returns ``STOP with failure'' for all edges $xy$ in a $P_3$ of $G$ {\bf then} STOP--$G$ has no d.i.m.\ {\bf else} STOP and return a d.i.m.\ of $G$. 
\end{itemize}

\end{algo}

\begin{theorem}\label{theo:DIMS124}
Algorithm $\ref{DIMS124fr}$ is correct and can be done in polynomial time. Thus, DIM can be solved in polynomial time for $S_{1,2,4}$-free graphs.
\end{theorem}

\noindent
{\bf Proof.} The correctness of the algorithm follows from the structural analysis of $S_{1,2,4}$-free graphs with a d.i.m.\ In particular: concerning Step (B), one can easily verify that if $G$ has a d.i.m.\ of one edge, then $G$ has no d.i.m.\ with more than one edge; concerning Step (C), one can refer to Observation \ref{obse:xy-in-P3}. 

\medskip

The time bound follows from the fact that Step (A) can be done in polynomial time (in particular the Edge C-Reduction can be done in polynomial time), Step (B) can be done in polynomial time, and Step (C) can be done in polynomial time by Theorem \ref{theo:procedureDIMxyS124}.
\qed 

\section{Conclusion}

It is still a widely open problem whether DIM can be solved in polynomial time for $S_{i,j,k}$-free graphs for any fixed $i,j,k$; for example, it is not clear how to solve it for $S_{1,3,4}$-free graphs or for $S_{2,2,4}$-free graphs but the approaches described here as well as in \cite{HerLozRieZamdeW2015} might be helpful.  

\medskip

\noindent
{\bf Acknowledgment.} The second author would like to witness that he just tries to pray a lot and is not able to do anything without that - ad laudem Domini.

\begin{footnotesize}

\end{footnotesize}

\end{document}